\documentclass[prd,amsmath,amssymb,amsfonts,nofootinbib,preprintnumbers,twocolumn,balancelastpage]{revtex4}
\usepackage{graphicx,slashed}
\usepackage{bbm}
\usepackage{color}

\usepackage{hyperref}
\hypersetup{
    pdfnewwindow=true,      
    colorlinks=true,        
    linkcolor=black,        
    citecolor=blue,         
    filecolor=blue,         
    urlcolor=blue           
}
\begin{document}
\title{\Large {\bf{The Electroweak Vacuum Angle}}}
\author{Pavel Fileviez P\'erez}
\affiliation{\vspace{0.15cm} \\  Particle and Astro-Particle Physics Division \\
Max-Planck Institute for Nuclear Physics {\rm{(MPIK)}} \\
Saupfercheckweg 1, 69117 Heidelberg, Germany}
\author{Hiren H. Patel}
\affiliation{\vspace{0.15cm} \\  Particle and Astro-Particle Physics Division \\
Max-Planck Institute for Nuclear Physics {\rm{(MPIK)}} \\
Saupfercheckweg 1, 69117 Heidelberg, Germany}

\begin{abstract}
We reveal a new source of CP-violation in the electroweak sector that is free of any experimental bounds, and we highlight the possible implications for baryogenesis.
\end{abstract}

\maketitle
Presence of CP-violating interactions is crucial to explain the observed baryon asymmetry of the universe \cite{Sakharov:1967dj}. The standard model features just two independent sources of CP violation: one as a phase in the Cabibbo-Kobayashi-Maskawa (CKM) matrix~\cite{CKM1,CKM2}, and the other in the form of the QCD vacuum angle, $\theta_\text{QCD}$~\cite{review}. In this letter, we address the possible existence of a new source of CP violation in the electroweak sector in direct analogy with that of the strong sector, which we call the electroweak vacuum angle $\theta_\text{EW}$.  

We demonstrate that in the standard model, the weak vacuum angle can be removed via the Adler-Bell-Jackiw (ABJ) anomaly \cite{Fujikawa:1979ay,Fujikawa:1980eg} thereby establishing that it has no physical consequences. We subsequently show that the necessary condition which renders $\theta_\text{EW}$ physical is the presence of simultaneous explicit violations of baryon and lepton conservation in the combination $B+L$.  This is in contrast with a previous study in Refs.~\cite{Anselm:1993uj,Anselm:1992yz}, where it is indicated that only explicit baryon number violation is sufficient.

At energies and intensities accessible to modern laboratories, the CP violating effect of $\theta_\text{EW}$ which proceeds via the electroweak instanton is severely suppressed, and offers no hope of measuring the parameter.  However, at the high temperatures present in the early universe, sphaleron rates are significantly enhanced, and the electroweak vacuum angle may provide a potentially large source of CP violation for baryogenesis.  Recently \cite{McLerran:2012mm}, the possibility that an axion associated with $\theta_\text{EW}$ might account for the cosmological constant was explored.


Before discussing the nature of the weak vacuum angle, we briefly review the situation concerning the QCD vacuum angle in the strong sector.  As is well-known the CP-violating term in the QCD Lagrangian
\begin{equation}\label{eq:QCDlag}
\mathcal{L}_\text{QCD}=-\frac{g^2_s \theta_\text{QCD}}{32\pi^2}G_{\mu\nu}^A \widetilde{G}^{\mu\nu\,A}
\end{equation}
is a total divergence.  Despite that, the term does not vanish on account of the existence of Yang-Mills instanton field configurations \cite{Belavin:1975fg,Jackiw:1976pf,Callan:1976je}.  But, the ABJ anomaly in the singlet axial current may be invoked to remove $\theta_\text{QCD}$ by a corresponding axial rotation $q_i \rightarrow  e^{-i\alpha\gamma_5}  q_i$ with the Lagrangian changing as
\begin{align}\label{eq:axialAnom}
\nonumber\delta\mathcal{L}_\text{QCD}&=\alpha\,\partial_\mu J_5^\mu\\
&=-\frac{n_q g_s^2 \alpha}{16\pi^2}G_{\mu\nu}^A \widetilde G^{\mu\nu\, A} + 2i\alpha\bar q M\gamma_5 q\,.
\end{align}
However, with non-zero quark masses the axial symmetry is explicitly broken.  Thus, one  accomplishes only in transferring $\theta_\text{QCD}$ from the topological term in Eq.~(\ref{eq:QCDlag}) into the quark mass matrix, as indicated by the second term in Eq.~(\ref{eq:axialAnom}).  Clearly, to eliminate $\theta_\text{QCD}$ one must search for a current whose divergence contains \emph{just} the anomaly with no other contributions; and then, make the corresponding field transformation.  But, because there are no such currents in QCD, it is impossible to eliminate $\theta_\text{QCD}$.

In the electroweak sector of the standard model the often-neglected topological term
\begin{equation}\label{eq:EWlag}
\mathcal{L}_\text{EW}=-\frac{g^2 \theta_\text{EW}}{32\pi^2}W_{\mu\nu}^a \widetilde{W}^{\mu\nu\,a}
\end{equation}
analogous to Eq.~(\ref{eq:QCDlag}) may be included.  While the Higgs condensate modifies the EW instanton tunneling rate \cite{'tHooft:1976fv}, the basic structure of topological vacua in the SU(2) sector remains unaltered \cite{Krasnikov:1979kz}.

As in QCD, $\theta_\text{EW}$ can be similarly adjusted if there are currents sensitive to the anomaly in weak isospin.  In the standard model, there are the well-known baryon and lepton currents.  Because these are classically conserved, under the corresponding vector U(1) transformations
\begin{equation}\label{eq:BLtransform}
Q \to e^{-i \alpha_\text{B}/3} Q, \qquad L \to e^{-i \alpha_{\text{L}}} L\,,
\end{equation}
the only change in the Lagrangian arises from the anomaly\footnote{For clarity, the contribution to the anomaly from the hypercharge gauge fields are omitted.}
\begin{align}
\nonumber\delta\mathcal{L}_\text{EW}&=\alpha_\text{B}\,\partial_\mu J^\mu_\text{B}+\alpha_\text{L}\,\partial_\mu J^\mu_\text{L}\\
&=-\frac{n_f g^2}{32\pi^2}(\alpha_\text{B}+\alpha_\text{L})W_{\mu\nu}^a \widetilde{W}^{\mu\nu\,a}\,,
\end{align}
amounting to a shift in the electroweak angle
\begin{equation}
\theta_\text{EW}\longrightarrow\theta_\text{EW}+n_f(\alpha_\text{B}+\alpha_\text{L})\,.
\end{equation}
Therefore, in the renormalizable standard model, any combination of transformation parameters satisfying $\alpha_\text{B}+\alpha_\text{L}=-\theta_\text{EW}/n_f$ can be chosen to remove the topological term from the Lagrangian. 

We now discuss the effect of explicit baryon and lepton number violation.  For concreteness, we shall treat the standard model as an effective field theory, and investigate the effect of adding the following operators to the Lagrangian
\begin{equation}
\mathcal{L}_{\text{BL}} = c_1 \frac{HH\ell\ell}{\Lambda} + c_2 \frac{qqq\ell}{\Lambda^2}  +  \ldots
\end{equation}
Written schematically, the first operator is responsible for Majorana neutrino masses and violates total lepton number;  the second operator mediates proton decay and violates baryon and lepton numbers.  This time, the four-divergence of the baryonic and leptonic currents acquire additional terms. 
The variation of the Lagrangian under the transformation given in Eq.~(\ref{eq:BLtransform}) is
\begin{multline}\label{eq:BLanom}
\delta\mathcal{L}=\alpha_\text{B}\,\partial_\mu J^\mu_\text{B}+\alpha_\text{L}\,\partial_\mu J^\mu_\text{L}=
-\frac{n_f g^2}{32\pi^2}(\alpha_\text{B}+\alpha_\text{L})W_{\mu\nu}^a \widetilde{W}^{\mu\nu\,a}
\\+i\alpha_\text{B}Q\frac{\partial \mathcal{L}_\text{BL}}{\partial Q}+i\alpha_\text{L}L\frac{\partial \mathcal{L}_\text{BL}}{\partial L}\,.
\end{multline}
This time, any choice for $\alpha_B$ or $\alpha_L$ would induce a phase in the couplings of the $B$ and $L$-violating interactions, as indicated by the additional terms in Eq.~(\ref{eq:BLanom}).  Because there are no other currents in the standard model whose divergence contains \emph{just} the SU(2) anomaly, it is impossible to remove $\theta_\text{EW}$, paralleling the situation in QCD.  

We emphasize that explicit non-conservation in only one of either baryon or lepton number is insufficient to render $\theta_\text{EW}$ physical because of the additional parameter of transformation offered by the other current which is still conserved.  Therefore, only in the case of simultaneous explicit violation of both baryon and lepton number, in the combination $B+L$, would $\theta_\text{EW}$ become a physical phase. This is in contrast with the results of Refs.~\cite{Anselm:1992yz,Anselm:1993uj} where they indicate baryon number violation is sufficient.

%
Here we illustrate how $\theta_\text{EW}$ enters into physical processes focusing, for simplicity, on the presence of one $B+L$ violating operator
\begin{equation}\label{eq:HighBL}
\frac{\lambda}{(\Lambda_\text{BL})^{14}}(q_L q_L q_L \ell_L)^3\,.
\end{equation}
Because of the large power suppression in $\Lambda_\text{BL}$ accompanying this operator, bounds on its presence is weak.  We may therefore consistently consider $\Lambda_\text{BL}\sim \text{TeV}$.  Such a large suppression may be  worrisome if we hope to see any effects of $\theta_\text{EW}$.  However, it should become clear by the following discussion that, for experiments at the intensity frontier (searches for electric dipole moments) and at the energy frontier (collider searches), an even bigger suppression renders this concern moot.

This large power suppression may be alleviated by considering the lower-dimension operator 
\begin{equation}\label{eq:GUTBL}
\frac{c_{ijkl}}{(\Lambda'_\text{BL})^2} u_R^i u_R^j d_R^k e_R^l
\end{equation}
derived from the GUT theory described in Ref.~\cite{Dorsner}.  The underlying group theoretic structure in the Yukawa coupling implies the coupling $c_{ijkl}$ must be antisymmetric under the generation indices $i\leftrightarrow j$, and consequently does not mediate proton decay.  Therefore the scale $\Lambda'_\text{BL}$ may be as low as $\sim \text{TeV}$.

CP-violating effects of the electroweak vacuum angle would appear in physical observables only through an interference of the non-perturbative instanton process with the perturbative operator in Eqs.~(\ref{eq:HighBL}) or (\ref{eq:GUTBL}).  Achieving interference with Eq.~(\ref{eq:GUTBL}), however, would require numerous chirality-flipping Higgs insertions, leading to additional suppression by powers of Yukawa couplings.

An illustrative example of the interference at work is the CP-conjugated pair of instanton-mediated $B+L$-violating processes at a hadron collider
\begin{gather}\label{eq:colliderProc}
\begin{aligned}
\text{I:}&\qquad q+q \longrightarrow 3\bar\ell + 7 \bar q\\
\text{II:}&\qquad\bar{q}+\bar{q} \longrightarrow 3\ell + 7 q\,.
\end{aligned}
\end{gather}
The amplitudes for these processes take the approximate form
\begin{align}
A(qq\rightarrow 3 \bar{\ell} + 7 \bar{q})&\sim\frac{i\lambda}{(\Lambda_\text{BL})^{14}}  +   \kappa \ e^{-\frac{8\pi^2}{g^2}-i\theta_\text{EW}}.\\
A(\bar{q}\bar{q}\rightarrow 3 \ell + 7 q)&\sim\frac{i\lambda}{(\Lambda_\text{BL})^{14}}  +   \kappa \ e^{-\frac{8\pi^2}{g^2}+i\theta_\text{EW}}.
\end{align}
The first term is the perturbative contribution, and the second term is the instanton mediated non-perturbative contribution containing dependence of the electroweak vacuum angle.  Interference among these two terms generates an asymmetry between the two processes in Eq.~(\ref{eq:colliderProc}) characterized by the asymmetry parameter
\begin{equation}
\mathcal{A}_\text{CP}\propto\text{Im}\,\big(\lambda e^{i\theta_\text{EW}}\big)\,.
\end{equation}
It is important to observe the basis independence of the asymmetry parameter: 
a vector transformation as in Eq.~(\ref{eq:BLtransform}) shifts $\theta_\text{EW}$ into the explicit $B+L$-violating coupling $\lambda$ 
so that it acquires a complex phase.  The value of the asymmetry parameter $\mathcal{A}_\text{CP}$, now the imaginary part of $\lambda$, remains unchanged.  

Unfortunately, any CP-violating effects derived from $\theta_\text{EW}$ is expected to be normalized by the enormous semi-classical suppression factor $e^{-8\pi^2/g^2}\sim 10^{-160}$ together with powers of inverse-cutoff scale $\Lambda_\text{BL}$.  Atomic and nuclear electric dipole moments are also suppressed in the same way.  Thus, the prospects for observing effects of $\theta_\text{EW}$ in laboratory settings is bleak\footnote{We point out, however, that by considering a more inclusive reaction involving the production of W and Higgs bosons, the exponent is corrected by the so-called `holy grail function,' and may mitigate the suppression.  See section XIII B in \cite{review} and references therein for further details.}.  However, this also implies the existence of a new source of CP-violation with no phenomenological constraints.  

This new source of CP-violation may play a significant role in early universe cosmology in connection to baryogenesis.  Following the above discussion we can say 
that there are two relevant scales: the electroweak scale $T_\text{EW}\sim100$ GeV and the scale of $B$ and $L$ violation $\Lambda_\text{BL}\sim \text{TeV}$. The exponential suppression disappears at temperatures above $T_\text{EW}$, and above $\Lambda_\text{BL}$ CP-violating reactions should proceed rapidly.  We expect the rate law for baryon number density $n_B$ to be modified as follows:
\begin{equation}
\frac{d n_B}{d t}+3 H n_B = - k(T)n_B + f(T,\,\mathcal{A}_\text{CP})\,,
\end{equation}
where $H$ is the Hubble constant, $k(T)$ is the CP-symmetric rate constant for $B$ violation, and $f(T,\,\mathcal{A}_\text{CP})$ is the effective CP-violating rate whose detailed form depends on the underlying UV-complete theory.  This avenue for baryogenesis appears to be promising, but certainly requires a detailed investigation. 

We conclude this letter with a brief summary.  We have pointed out a potentially new source of CP-violation in the electroweak sector, and determined the conditions on the underlying Lagrangian that renders it physical.  We have provided an illustrative example of how it affects physical processes.  Finally, we highlighted its role in early universe cosmology.

{\textit{Acknowledgments:}}
We would like to thank Mark B. Wise for scrutinizing an earlier draft of this paper, and Michael J. Ramsey-Musolf for critical discussions regarding low energy phenomenology.  We especially thank Andrew J. Long for extensive clarifying conversations in connection to early universe cosmology.



\end{document}